\documentclass[useAMS,usenatbib]{mn2e}
\usepackage{graphicx}
\usepackage{pslatex}
\usepackage{natbib}
\usepackage{txfonts}

\newcommand{\vv}[1]{\bmath{#1}}

\begin{document}

\title[YORP effect with anisotropic radiation]{YORP effect with anisotropic radiation}

\author[S. Breiter and D. Vokrouhlick\'{y}]{S. Breiter$^{1}$\thanks{E-mail:
breiter@amu.edu.pl} and D. Vokrouhlick\'{y}$^{2}$\thanks{E-mail: vokrouhl@cesnet.cz} \\
$^{1}$Astronomical Observatory, A. Mickiewicz University, Sloneczna 36, PL60-286 Pozna\'{n}, Poland\\
$^{2}$Institute of Astronomy, Charles University, V Hole\v{s}ovi\v{c}k\'{a}ch 2, 18000 Prague 8, Czech Republic.}

\date{}

\pagerange{\pageref{firstpage}--\pageref{lastpage}} \pubyear{2010}

\maketitle
\label{firstpage}

   \begin{abstract}
    The influence of optical scattering and thermal radiation models an the Yarkovsky-O'Keefe-Radzievskii-Paddack (YORP) effect
    is studied. Lambertian formulation is compared with Hapke scattering and emission laws and Lommel-Seeliger reflection.
    Although the form of reflectivity function strongly influences mean torques due to scattering or thermal radiation alone,
    their combined contribution to the rotation period YORP is not much different from the standard Lambertian values.
    For higher albedo values the differences between the Hapke and Lambert models become significant for the YORP in attitude.
   \end{abstract}
   \begin{keywords}
   {radiation mechanisms: thermal---methods: numerical---celestial
   mechanics---minor planets, asteroids: general}
   \end{keywords}

\section{Introduction}

The importance of radiation recoil forces on both orbital motion and rotation of minor bodies in the Solar System has been commonly appreciated
over the last decade. The Yarkovsky effect, caused by lagged thermal radiation from the surface of a spinning body (directly detected in the
orbital motion of 6489 Golevka \citep{ChOV:2003} and 1992~BF \citep{VChM:2008}), has occurred to be the key to the proper understanding of asteroid
long-term dynamics. Since the paper of \citet{Rub:00}, the influence of torques due to radiation recoil is known as the YORP (Yarkovsky-O'Keefe-Radzievskii-Paddack)
effect, acknowledging the works of \citet{Yark:01}, \citet{Radz:54}, \citet{Pad:69}, and \citet{OK:76}.
Unlike the orbital Yarkovsky effect, YORP involves both the scattering of incident light and its thermal re-radiation, and it occurs
even for objects with zero conductivity. Direct detections of YORP effect in the rotation of asteroids 54509~YORP \citep{Low:07,Tay:07},
1862~Apollo \citep{Kaasa:07}, 1620~Geographos \citep{DVKH:2008}, and 3103~Eger \citep{DVP:2009DPS}, has proved the existence of YORP.
On the other hand, however, the agreement between the observed and modeled values in each case can be qualified as merely having a similar
order of magnitude and all present YORP models are still simplified and incomplete. What is worse, the failure to detect a theoretically
predicted YORP effect for 25143 Itokawa \citep{DVK:08} has helped to realize an extreme sensitivity of these simplified models to the fine
details of shape, centre of mass location and spin axis orientation in the body frame \citep{DVK:08,SG:08,Statler:09,BBCOV:09}.

Most of these models assume that both scattering and thermal radiation is Lambertian, i.e. a photon can be emitted or scattered with equal probability
in any direction, hence the exiting flux is proportional to the cosine of zenith distance according to the projected area of the radiating/scattering
surface element. Although \citet{BMV:07} mentioned a more general scattering model, their work on YORP for spheroids was based on the Lambertian
assumption. \citet{Sch:07} added a specular reflection to the Lambertian scattering, but such an improvement has no effect on the mean torque, as observed
by \citet{Rub:00} and proved by \citet{NV:08a} and \citet{RP:2010}. \citet{Statler:09} made a step further, using a hemispheric albedo derived from the scattering
model of \citet{Hapke:2002} instead of the usual Bond albedo \citep{VB:2001}. Moreover, the TACO model of Statler for the first time incorporates
the important observation that photons bouncing between various surface elements do not produce the net torque until they they finally exit into
outer space.

The main objective of the present work is to include a non-Lambertian scattering and radiation into the recent YORP model of \citet{BBC:2010} and judge
the significance of this improvement. Roughly speaking, a departure from the Lambertian model is essentially caused by interreflections and occlusions.
Both phenomena occur at various levels of resolution and one has to be careful about this issue. Out of the two principal scattering models for asteroids
surface developed by \citet{LumBo:1981} and Hapke \citep{Hapke:1981,HW:1981,Hapke:1984,Hapke:1986,Hapke:2002,Hapke:2008} we have chosen the latter.
However, both models were created to interpret photometric observations; as such, they attempt to include phenomena happening at various resolution
levels that merge in the final integrated brightness. In these circumstances, our present paper focuses on accounting for the regolith grain size
($ < 1~\mathrm{mm}$) scale phenomena described by an appropriate part of the Hapke reflectance and emissivity models. It means ignoring
the macroscopic roughness corrections and shape-dependent beaming factors. Interreflections occurring between larger surface fragments require
a different approach and will be discussed in another article, whereas the large scale occlusions (shadowing)
are already incorporated in most of existing YORP models. To a large extent the present contribution
has been motivated by the problem of YORP effect on a high albedo asteroid 3103~Eger, where only a convex shape model is available,
so larger scale interreflections have to be neglected anyway.

We decided to present a detailed derivation of the radiation recoil force and associated YORP torque using the terminology of modern radiometry
instead of  traditional astrophysical framework dating back to \citet{Chan:1950}. In this respect we owe much to the collection
of Max Fairbairn essays available on-line thanks to J.B. Tatum (\verb"http://orca.phys.uvic.ca/~tatum/plphot.html").

\section{Scattering and radiation}
\label{scatra}

\subsection{Irradiance}

Consider an infinitesimal element $\mathrm{d}S$ of an atmosphereless celestial body surface. Further, we call $\mathrm{d}S$ \emph{a physical surface},
to distinguish it from \emph{a normal surface} i.e. an infinitesimal surface perpendicular to some specified direction of incident or emitted radiation.

Local solar frame (LSF) will be defined with the origin at the centre of $\mathrm{d}S$, with
$x$-axis pointing to the intersection of the meridian passing through the Sun and the horizon plane, $z$-axis directed along the outward normal to
the physical surface (i.e. to the local zenith), and $y$-axis completing the right-handed orthogonal system.
Then the unit vector directed to the Sun has a simple form
\begin{equation}\label{ns}
    \vv{s} = \left(
                     \begin{array}{c}
                       s_\odot\\
                       0 \\
                       \mu_\odot \\
                     \end{array}
                   \right),
\end{equation}
depending only on the solar zenith distance   through its cosine  $\mu_\odot$ and sine
\begin{equation}\label{sodot}
    s_\odot =  \sqrt{1-\mu_\odot^2}.
\end{equation}

If the Sun is located at the distance $R_\odot$, the \emph{collimated radiation flux density} (power per normal area perpendicular to $\vv{s}$)
arriving from the point $R_\odot\,\vv{s}$ is
\begin{equation}\label{flux:dens}
     J = J_0 \, \left(\frac{R_0}{R_\odot}\right)^2,
\end{equation}
where the Solar constant $J_0 \approx 1366~\mathrm{W\,m^{-2}}$ is defined for nominal distance $R_0 = 1~\mathrm{au}$.
\emph{Irradiance} or \emph{insolation} $E$ of an arbitrarily oriented surface element is the ratio of the incident power flux $\mathrm{d}\Phi_\mathrm{i}$ to
the physical area $\mathrm{d}S$. Accounting for the area projection factor $\vv{s}\cdot \vv{n} = \mu_\odot$, where
$\vv{n}$ is the unit vector directed to zenith, we can write
\begin{equation}\label{irrad:0}
    E(\vv{s}) = \frac{\mathrm{d}\Phi_\mathrm{i}}{\mathrm{d}S} =  \nu J  \mu_\odot.
\end{equation}
The visibility function $\nu$ is either 1, when the Sun is visible at $\mathrm{d}S$, or 0, when the Sun is occluded.

\subsection{Bidirectional scattering}

Incident radiant power $\mathrm{d}\Phi_\mathrm{i} = E\,\mathrm{d}S$ is partially absorbed (converted into heat) and partially reflected
in various directions $\vv{o}$
\begin{equation}\label{o}
    \vv{o} = \left(
                     \begin{array}{c}
                       \sqrt{1-\mu^2} \cos{\phi} \\
                       \sqrt{1-\mu^2} \sin{\phi} \\
                       \mu \\
                     \end{array}
                   \right),
\end{equation}
where $\mu$ is the cosine of the zenith distance and $\phi$ is the azimuth angle in LSF.
The cosine of the phase angle between $\vv{s}$ and $\vv{o}$ will be designated $\mu'$, and defined as the scalar product
\begin{equation}\label{muprim}
    \mu' = \vv{s} \cdot \vv{o} = s_\odot  \sqrt{1-\mu^2} \cos{\phi} + \mu_\odot \mu.
\end{equation}

The power flux $\mathrm{d}^2\Phi_\mathrm{r}$ scattered from $\mathrm{d}S$ into a solid angle $\mathrm{d}\Omega$ in direction $\vv{o}$ is
described by \emph{reflected radiance}  $L_\mathrm{r}$
\begin{equation}\label{radiance}
   L_\mathrm{r}(\vv{o} ) =  \frac{\mathrm{d}^2\Phi_\mathrm{r}}{\mu\,\mathrm{d}S\,\mathrm{d}\Omega},
\end{equation}
where $\mu\,\mathrm{d}S$ is the normal surface  perpendicular to $\vv{o}$.
Writing $L_\mathrm{r}(\vv{o})$ we should bear in mind an implicit dependence on the Sun direction $\vv{s}$, because the reflected power depends on
the incident flux from the Sun as well. This dependence becomes more explicit, when we introduce a bidirectional reflectance distribution function (BRDF) $f_\mathrm{r}$
defined as the ratio of the radiance $L_\mathrm{r}$ reflected in the direction $\vv{o}$ to
the irradiance $E$ from the energy source located in the direction $\vv{s}$
\begin{equation}\label{BRDF}
  f_\mathrm{r}(\vv{s},\,\vv{o})  =  \frac{ L_\mathrm{r}(\vv{o}) }{   E(\vv{s})}.
\end{equation}
Although a bidirectional reflectance (BDR) function~$\rho$
\begin{equation}\label{rho}
  \rho(\vv{s},\vv{o}) = \mu_\odot \, f_\mathrm{r}(\vv{s},\vv{o}),
\end{equation}
seems to be more common in planetary photometry than BRDF, we choose $f_\mathrm{r}$ as a more convenient quantity offering, for example, the reciprocity relation
$f_\mathrm{r}(\vv{s},\,\vv{o}) = f_\mathrm{r}(\vv{o},\,\vv{s})$.
Using Eq.~(\ref{BRDF}) we can express the reflected radiance as
\begin{equation}\label{Lr}
      L_\mathrm{r}(\vv{o}) =  f_\mathrm{r}(\vv{s},\vv{o})\, E(\vv{s}) = \nu\,f_\mathrm{r}(\vv{s},\vv{o})\,\mu_\odot\,J.
\end{equation}
~\\

Recalling for the reference a traditional, Lambertian BRDF with albedo $A$
\begin{equation}\label{Lamb}
    f_\mathrm{L} = \frac{A}{\pi},
\end{equation}
we adopt the anisotropic BRDF proposed by Hapke, namely its version from \citep{Hapke:2002} without macroscopic roughness effects
\begin{equation}\label{BDRF:Hap}
    f_\mathrm{r}(\vv{s},\vv{o}) =   \frac{w}{4\pi\,(\mu_\odot+\mu)} \left[ \left( 1+ B  \right)  P  + H(\mu_\odot) H(\mu) - 1\right],
\end{equation}
where the Henyey-Greenstein particle phase function is
\begin{equation}\label{p}
    P = \left(1-g^2\right) \,\left[ 1 + 2\,g\,\mu' + g^2 \right]^{-\frac{3}{2}},
\end{equation}
and the opposition surge function $B$ is defined as
\begin{equation}\label{B}
 B = B_0\,\left[ 1+\frac{1}{h} \sqrt{\frac{1+\mu'}{1-\mu'}} \, \right]^{-1},
\end{equation}
with
\begin{equation}\label{B0}
 B_0 = \frac{S_0}{w} \frac{(1+g)^2}{(1-g)}.
\end{equation}

The Chandrasekhar multiple scattering function $H$ is defined in terms of an integral equation \citep{Chan:1950}, but we use the
explicit second order approximation given by \citet{Hapke:2002}
\begin{equation}\label{H}
    H(x) = \left[ 1 - w x \left( r_0 + \frac{1-2 r_0 \,x}{2} \ln{\frac{1+x}{x}} \right)\right]^{-1},
\end{equation}
where
\begin{equation}\label{gamma}
    r_0  = \frac{1 - \sqrt{1-w}}{1 + \sqrt{1-w}}.
\end{equation}
Thus, apart from the incoming and scattered flux directions $\vv{s}$ and $\vv{o}$, the Hapke BRDF depends
on four physical parameters of the surface: single scattering albedo $w$, regolith compaction parameter $h$, the opposition surge amplitude $S_0$
(sometimes replaced by $B_0$), and the asymmetry factor of the Henyey-Greenstein function $g$. A more recent version of reflectance was
proposed by  \citet{Hapke:2008};
the modification amounts to adding the dependence on porosity as a multiplicative factor in $f_\mathrm{r}$ and a divisor in the argument of $H$.
This modification is easy to implement, but we suspend its use until controversies concerning the dependence of the opposition effect on porosity
are resolved \citep{Hapke:2008}.\\

The total power flux $\Phi_\mathrm{e}$ emitted into the hemisphere
$$\Omega_+ = \left\{(\mu, \phi): ~0 \leq  \mu \leq  1, ~ 0 \leq  \phi < 2 \pi \right\},$$
divided by the emitting physical area is termed \emph{radiant exitance} $M$
\begin{equation}\label{exit}
    M  = \frac{\mathrm{d}\Phi_\mathrm{e}}{\mathrm{d}S}.
\end{equation}
Recalling the definition (\ref{radiance}), we find for the exitance due to scattering
\begin{equation}\label{exit:1}
    M_\mathrm{r}  = \int_{\Omega_+} \frac{\mathrm{d}^2\Phi_\mathrm{r}}{\mathrm{d}S\,\mathrm{d}\Omega}\,\mathrm{d}\Omega
    = \int_{\Omega_+} L_\mathrm{r}(\vv{o},\,\vv{s})\,\mu\,\mathrm{d}\Omega.
\end{equation}
But, according to Eq.~(\ref{BRDF}), radiance is related to irradiance by the BRDF $f_\mathrm{r}$, hence
\begin{equation}\label{exit:2}
    M_\mathrm{r}(\vv{s})  = E(\vv{s})\, \int_{\Omega_+} f_\mathrm{r}(\vv{s},\vv{o})\,\mu\,\mathrm{d}\Omega,
\end{equation}
and the dependence on the Sun location $\vv{s}$ appears explicitly.

At this point we can introduce the notion of hemispheric albedo $A_h$ as the ratio
\begin{equation} \label{Ah:def}
    A_\mathrm{h}(\mu_\odot) = \frac{M_\mathrm{r}}{E(\vv{s})}.
\end{equation}
Combining Eqs.~(\ref{exit:2}) and (\ref{Ah:def}) we see that
\begin{equation}\label{HA}
    A_\mathrm{h}(\mu_\odot) =   \int_{\Omega_+}f_\mathrm{r}(\vv{s},\,\vv{o})\,\mu\, \mathrm{d}\Omega =
     \int_0^{2\pi} \mathrm{d}\phi \int_0^1 f_\mathrm{r}(\vv{s},\,\vv{o})\,\mu\, \mathrm{d}\mu,
\end{equation}
and Eq.~(\ref{exit:2}) is simplified to
\begin{equation}\label{exit:2a}
    M_\mathrm{r}(\vv{s})  = E(\vv{s})\,A_\mathrm{h}(\mu_\odot).
\end{equation}

For a given set of Hapke parameters the integral (\ref{HA}) can be evaluated  numerically on a sufficiently dense set of $\mu_\odot$ values,
allowing to construct an appropriate approximating function. Using the least squares adjustment we construct
\begin{equation}\label{mAh}
    \mu_\odot\, A_\mathrm{h}(\mu_\odot) \approx  A_\mathrm{B} \mu_\odot + \alpha_1\,\mu_\odot\,\frac{1 - \alpha_2 \mu_\odot}{1+\alpha_3 \mu_\odot},
\end{equation}
where the Bond albedo  $A_\mathrm{B}$, defined as
\begin{equation}\label{AB}
    A_\mathrm{B} = \frac{1}{\pi} \int_{\Omega_+}A_\mathrm{h}(\mu_\odot)\,\mu_\odot \mathrm{d}\Omega_\odot =
    2   \int_0^1 A_\mathrm{h}(\mu_\odot)\,\mu_\odot \mathrm{d}\mu_\odot,
\end{equation}
is the mean slope of the product $\mu_\odot A_\mathrm{h}$, and the coefficients $\alpha_i$ of a simple rational approximation
describe the deviation from the linear model. We focus on the properties of $\mu_\odot A_\mathrm{h}$, because in next sections
the hemispheric albedo will always appear multiplied by the cosine of the Sun's zenith distance. Note, that the adjustment of
$A_\mathrm{h}(\mu_\odot)$ leads to different values of $\alpha_i$, degrading the quality of approximation of the product $\mu_\odot\, A_\mathrm{h}$.

\subsection{Geometric albedo}

Although the geometric (or physical) albedo  is not directly involved in the computation of radiation recoil force,
we need it to select an appropriate value of $w$ for the Hapke model, since usually the observations provide for an asteroid
only the geometric albedo and the spectral type.

Let us begin with the notion of \emph{intensity} $I$. In contrast to the previously discussed quantities, intensity refers to the power $\mathrm{d}\Phi$
emitted from the whole body surface (not only from an infinitesimal $\mathrm{d}S$) in some direction $\vv{\hat{q}}$, divided by the solid angle $\mathrm{d}\Omega$
centered at $\vv{\hat{q}}$
\begin{equation}\label{intens}
    I(\vv{\hat{q}}) = \frac{\mathrm{d}\Phi}{\mathrm{d}\Omega}.
\end{equation}
The geometric albedo $p$ is the ratio of observed intensity of some presumably spherical object to the intensity of Lambertian disk
with the same diameter as the  assumed sphere -- both observed from the direction to the Sun, i.e. with zero phase angle.
This leads to the integral definition
\begin{equation}\label{geom}
    p = 2\pi \int_0^1 \mu_\odot^2 \,f_\mathrm{r}(\vv{s},\, \vv{s})\,\mathrm{d}\mu_\odot.
\end{equation}

\citet{VV:95} provide expressions that allow to compute Hapke parameters $h$, $B_0$, $g$ for various
spectral types as functions of a given geometric albedo $p$ and the mean slope parameter $G$ of the IAU two-parameter magnitude system \citep{BHD:1989}.

\subsection{Directional thermal emission}

The energy leaving a surface element $\mathrm{d}S$ does not consist only of scattered radiation. If the element has temperature $T >0$, it also emits
thermal radiation. Radiant exitance $M_\mathrm{b}$ through $\Omega_+$ for a black body is given by the Stefan-Boltzmann law
\begin{equation}\label{bb}
   M_\mathrm{b} = \frac{\mathrm{d}\Phi_\mathrm{b}}{\mathrm{d}S} = \sigma\,T^4,
\end{equation}
where $\sigma = 5.67 \times 10^{-8}~\mathrm{W\,m^{-2}\,K^{-4}}$, is the Stefan-Boltzmann constant, and $\Phi_\mathrm{b}$ is the black body value
of a more general  thermal radiation power flux $\Phi_\mathrm{t}$.
The point black body  radiation is by definition isotropic, whereas a black body surface radiation is Lambertian,
 so the associated radiance $L_\mathrm{b}(\vv{o})$ in the direction $\vv{o}$ is obtained
from the general definition of a \emph{thermally emitted radiance} $L_\mathrm{t}$ analogous to (\ref{Lr})
\begin{equation}\label{radiance:t}
   L_\mathrm{t}(\vv{o}) =  \frac{\mathrm{d}^2\Phi_\mathrm{t}}{\mu\,\mathrm{d}S\,\mathrm{d}\Omega},
\end{equation}
dividing the exitance $M_\mathrm{b}$ by the `$\mu$-averaged' solid angle of a hemisphere $\pi$
\begin{equation}\label{LB}
  L_\mathrm{b}  = \frac{M_\mathrm{b}}{\pi}.
\end{equation}
Indeed, using (\ref{LB}) and the definition of exitance, we verify that
\begin{equation}\label{test}
     \int_{\Omega_+}  L_\mathrm{b}\,\mu \,\mathrm{d}\Omega = M_\mathrm{b}.
\end{equation}

\emph{Hemispheric emissivity} $\epsilon_\mathrm{h}$ is defined as the ratio of actual thermal exitance $M_\mathrm{t}$ to the black body exitance $M_\mathrm{b}$
\begin{equation}\label{eps:t}
 \epsilon_\mathrm{h} = \frac{M_\mathrm{t}}{M_\mathrm{b}}.
\end{equation}
This global quantity should not be confused with a \emph{directional emissivity} $\epsilon(\vv{o})$, defined as the ratio of radiances
\begin{equation}\label{epsdef:0}
    \epsilon(\vv{o}) = \frac{ L_\mathrm{t}(\vv{o})}{ L_\mathrm{b}(\vv{o})}.
\end{equation}
Directional emissivity plays the role similar to that of BRDF in scattering, although their definitions essentially differ: the former is dimensionless ratio of two radiances,
while the latter (with dimension $\mathrm{sr}^{-1}$) is the ratio of radiance to irradiance. So, the thermally emitted radiance in the direction of
$\vv{o}$ can be expressed as
\begin{equation}\label{Lt}
 L_\mathrm{t}(\vv{o}) =    \epsilon(\vv{o})\,L_\mathrm{b}(\vv{o}) =  \frac{\epsilon(\vv{o})}{\pi}\,\sigma\,T^4.
\end{equation}
~\\

In present study we adopt the directional emissivity function of Hapke  \citep{Hapke:93,Lag:96}
\begin{equation}\label{eps}
    \epsilon(\mu) = \sqrt{1-w}\,H(\mu),
\end{equation}
where $w$ is Hapke's single scattering albedo, and the Chandrasekhar function $H$ is given by Eq.~(\ref{H}). Thus the emitted radiance is
\begin{equation}\label{radiance:eps}
   L_\mathrm{t}(\vv{o}) =  \epsilon(\mu) \, \frac{\sigma\,T^{4}}{\pi} = \frac{\sqrt{1-w}}{\pi}\,H(\mu)\,\sigma\,T^4.
\end{equation}

The exitance $M_\mathrm{t}$ resulting from Eq.~(\ref{radiance:eps}) is
\begin{equation}\label{Mt}
     M_\mathrm{t}  = \int_{\Omega_+} L_\mathrm{t}(\vv{o})\,\mu\,\mathrm{d}\Omega.
\end{equation}
Confronting it with the primary definition of $\epsilon_\mathrm{h}$ (\ref{eps:t}), we can use the relation
\begin{equation}\label{exit:4}
    M_\mathrm{t}   = \epsilon_\mathrm{h}\,\sigma \,T^4,
\end{equation}
which leads to the integral expression of hemispheric emissivity
\begin{equation} \label{epst}
\epsilon_\mathrm{h} =   \int_{\Omega_+} \frac{\epsilon(\mu)\,\mu}{\pi}\,\mathrm{d}\Omega = 2 \,\int_{0}^{1} \epsilon(\mu)\,\mu\,\mathrm{d}\mu,
\end{equation}
evaluating to a single number for a given set of Hapke parameters.

The infrared radiation of asteroids is often related with the notion of beaming effect, empirically accounted for by the beaming factor $\eta$ \citep{LS:89,Lag:96}.
We do not introduce the beaming factor in our model  for a number of reasons: i) the part of beaming that depends on grain size scale radiation transfer
should be present in the emissivity function of Hapke, ii) the contribution of thermal lag to the beaming factor is present in the surface temperature model with conductivity,
iii) larger scale radiation exchange contribution \citep{Lag:98} will be included in future extensions of our model together with optical interreflections.

\subsection{Energy balance}

Conservation of energy implies, that the total power scattered, thermally re-emitted, and conducted inside the body, should be equal to the incident power flux $\Phi_\mathrm{i}$.
In terms of power density (per physical surface) it means that
\begin{equation}\label{ener:bal}
    E(\vv{s}) = M_\mathrm{r}(\vv{s})+M_\mathrm{t}-Q,
\end{equation}
i.e. irradiance E is equal to the sum of total radiant exitance $M$ and of the conducted heat flux density  $(-Q)$.
Given a nonzero surface conductivity $K$, we have
\begin{equation}\label{bc:1}
 Q =  - K\,\vv{n} \cdot \nabla T,
\end{equation}
and then
\begin{equation}\label{ener:1}
    E(\vv{s}) =   A_\mathrm{h}(\mu_\odot)\,E(\vv{s})  + \epsilon_\mathrm{h}\,\sigma \,T^4 + K\,\vv{n} \cdot \nabla T,
\end{equation}
or
\begin{equation}\label{ener:2}
    \epsilon_\mathrm{h}\,\sigma \,T^4 =  \nu\,\mu_\odot \,\left(1 - A_\mathrm{h}(\mu_\odot) \right)\,J + Q.
\end{equation}
If $K=0$, Eq.~(\ref{ener:2}) directly provides the surface temperature, generalizing the usual Lambertian Rubincam approximation of the YORP effect
to the Hapke reflectance/emissivity model.
With $K \neq 0$,  Eq.~(\ref{ener:2}) serves as a boundary condition for the heat conduction problem.

\section{Radiation recoil force and torque}

\subsection{Force expression}

Photon flux, leaving $\mathrm{d}S$ in direction $\vv{o}$, carries energy and momentum (energy divided by the velocity of light $c$),
inducing the recoil force $\vv{F}$ equal to the time derivative of momentum and directed opposite to $\vv{o}$.
The force can be easily expressed in terms of emitted radiance, provided we introduce a \emph{radiance vector}
\begin{equation}\label{radiance:v}
\vv{L}(\vv{o}) = L(\vv{o})\,\vv{o},
\end{equation}
where $L$ is the sum of scattered $L_\mathrm{r}$ and thermal $L_\mathrm{t}$. The definition of radiance implies that the force density
in the direction $\vv{o}$ per physical area and solid angle is
\begin{equation}\label{dF:0}
    \frac{\mathrm{d}^2\vv{F}}{\mathrm{d}S\,\mathrm{d}\Omega} =
    - \frac{\mathrm{d}^2(\Phi_\mathrm{r}+\Phi_\mathrm{t})}{\mathrm{d}S\,\mathrm{d}\Omega}\,\frac{\vv{o}}{c}
     =  - \frac{\mu}{c}\,\vv{L}(\vv{o}).
\end{equation}
Integrating over the hemisphere $\Omega_+$, we find the net force per physical area
\begin{equation}\label{dF:1}
    \frac{\mathrm{d}\vv{F}}{\mathrm{d}S} =
   - \frac{1}{c} \int_{\Omega_+} \mu\,\vv{L}(\vv{o})\,\mathrm{d}\Omega.
\end{equation}
Substituting the expressions (\ref{Lr}) and (\ref{Lt}), we have
\begin{equation}\label{dF:2}
    \frac{\mathrm{d}\vv{F}}{\mathrm{d}S} =
   - \frac{1}{c} \int_{\Omega_+} \mu\,\left(    f_\mathrm{r}(\vv{s},\vv{o})\, E(\vv{s}) +
   \frac{\epsilon(\mu)}{\pi}\,\sigma\,T^4 \right)\,\vv{o}\,\mathrm{d}\Omega,
\end{equation}
or, observing the independence of directional emissivity on azimuth
\begin{equation}\label{dF:3}
    \frac{\mathrm{d}\vv{F}}{\mathrm{d}S} =
   - \frac{E(\vv{s})}{c} \int_{\Omega_+} \mu\,  f_\mathrm{r}(\vv{s},\vv{o}) \,\vv{o}\,\mathrm{d}\Omega
 - \frac{2\,\sigma\,T^4\,\vv{n}}{c} \int_0^1 \mu^2\,
    \epsilon(\mu) \,\mathrm{d}\mu.
\end{equation}

Using LSF we can conveniently decompose the force density into the sum of two perpendicular components along the axes $z$ and $x$,
i.e. along the surface normal $\vv{n}$ and the unit vector $\vv{m}$
\begin{equation}\label{nm}
    \vv{n} = \left(
                     \begin{array}{c}
                       0 \\
                       0 \\
                       1 \\
                     \end{array}
                   \right),
    \qquad
\vv{m} = \left(
                     \begin{array}{c}
                       1 \\
                       0 \\
                       0 \\
                     \end{array}
                   \right).
\end{equation}
In an arbitrary reference frame we can compute $\vv{m}$ as
\begin{equation}\label{marb}
    \vv{m} = \left( \vv{s} - \mu_\odot \, \vv{n}\right) \, s_\odot^{-1},
\end{equation}
taking $\mu_\odot = \vv{s} \cdot \vv{n}$.

Splitting the first integrand in Eq.~(\ref{dF:3}) into a sum
\begin{eqnarray}
   \mu\,  f_\mathrm{r}(\vv{s},\vv{o}) \,\vv{o} &=& f_\mathrm{r}\, \mu^2\,\vv{n}
   + f_\mathrm{r} \,\mu\,\sqrt{1-\mu^2} \cos{\phi}\,\vv{m} \nonumber \\
    & & + f_\mathrm{r} \,\mu\,\sqrt{1-\mu^2} \sin{\phi}\,\left(\vv{n} \times \vv{m}\right),
    \label{integr}
\end{eqnarray}
and recalling that  $ f_\mathrm{r}(\vv{s},\vv{o})$ is an even function of azimuth $\phi$, hence the last term
in Eq.~(\ref{integr}) is odd and its integral over $\Omega_+$ does vanish, we introduce two auxiliary functions
\begin{eqnarray}
  I_1(\mu_\odot) &=& \int_0^{2\pi} \mathrm{d}\phi \int_0^1 \mu^2 \,f_\mathrm{r}(\vv{s},\vv{o})\, \mathrm{d}\mu, \\
  I_2(\mu_\odot) &=& \int_0^{2\pi} \mathrm{d}\phi  \int_0^1 \mu \sqrt{1-\mu^2} \cos{\phi} \,f_\mathrm{r}(\vv{s},\vv{o})\, \mathrm{d}\mu,
\end{eqnarray}
as well as a coefficient
\begin{equation}\label{I3}
     I_3  =   2 \int_0^1 \mu^2\,\epsilon(\mu)  \,\mathrm{d}\mu,
\end{equation}
that allow to rewrite the formula (\ref{dF:3}) as
\begin{equation}\label{dF:4}
    \frac{\mathrm{d}\vv{F}}{\mathrm{d}S} =
   - \frac{E(\vv{s})}{c}  \left[ I_1(\mu_\odot) \,\vv{n} + I_2(\mu_\odot) \,\vv{m} \right]
 - \frac{I_3\,\sigma\,T^4}{c}\,\vv{n}.
\end{equation}

Substituting the boundary conditions (\ref{ener:1}) into (\ref{dF:4}) we remove the explicit dependence on $T^4$, obtaining
\begin{equation}\label{dF:5}
  \frac{\mathrm{d}\vv{F}}{\mathrm{d}S} =
    - \frac{E(\vv{s})}{c}\,   \left[ \left(I_1(\mu_\odot) + I_3\,\frac{1-A_\mathrm{h}(\mu_\odot)}{\epsilon_\mathrm{h}}\right) \,\vv{n}
     + I_2(\mu_\odot) \,\vv{m} \right]
     - \frac{Q\,I_3}{c\,\epsilon_\mathrm{h}}\,\vv{n}.
\end{equation}
However, we prefer to rearrange the force expression into a more comprehensive form
\begin{equation}\label{dF:fin}
  \frac{\mathrm{d}\vv{F}}{\mathrm{d}S} = - \frac{2}{3}\,\frac{1+\xi}{c} \left[ \nu J \mu_\odot  + Q \right]\,\vv{n}
    + \frac{\nu J}{c}\, \left[  X_1  \,\vv{n}
  -  X_2  \,\vv{m} \right],
\end{equation}
where functions arguments have been omitted for the sake of brevity.

\begin{table}
\caption{Sample Hapke parameters and related quantities}

\begin{tabular}{|c|c|c|}
  \hline
  & $p = 0.6$, type E & $p=0.1$, type S\\
  \hline
  $w$ & $0.856$ & $0.139$\\
  $h$ & $0.044$ & $0.049$  \\
  $B_0$ & $0.8576$ & $1.5407$ \\
  $g$ & $ -0.2459$  & $-0.2593$ \\
  \hline
 $A_\mathrm{B}$ & $0.47542$ & $0.046712$  \\
  $\alpha_1$ & $0.14550$ & $0.031021$  \\
  $\alpha_2$ & $1.5880$ &  $1.6968$  \\
  $\alpha_3$ & $0.9741$ &  $3.6800$ \\
  \hline
 $\xi_{10}$ & $0.09590$ &  $0.01212$ \\
 $\xi_{11}$ & $-0.03374$ & $0.3383$ \\
 $\xi_{12}$ & $2.1268$ & $1.4786$ \\
 $\xi_{20}$ & $0.032156$ & $0.006259$  \\
 $\xi_{21}$ & $0.42458$ & $0.16050$  \\
 \hline
 $\xi$ & $0.03024$ & $0.00255$ \\
 $\epsilon_\mathrm{h}$ & $0.58832$ & $0.96905$ \\
 \hline
\end{tabular}
\label{tab:hap}
\end{table}

\begin{figure}
 \begin{center}
  \includegraphics[width=8cm]{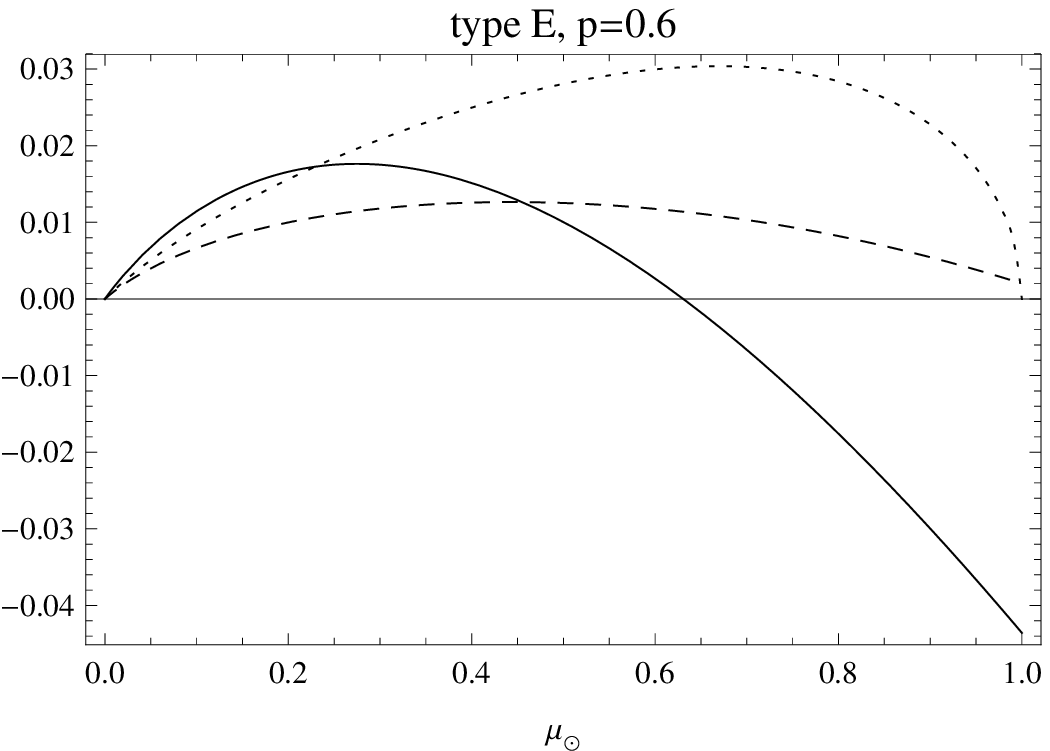}\\ \vskip0.5cm
  \includegraphics[width=8cm]{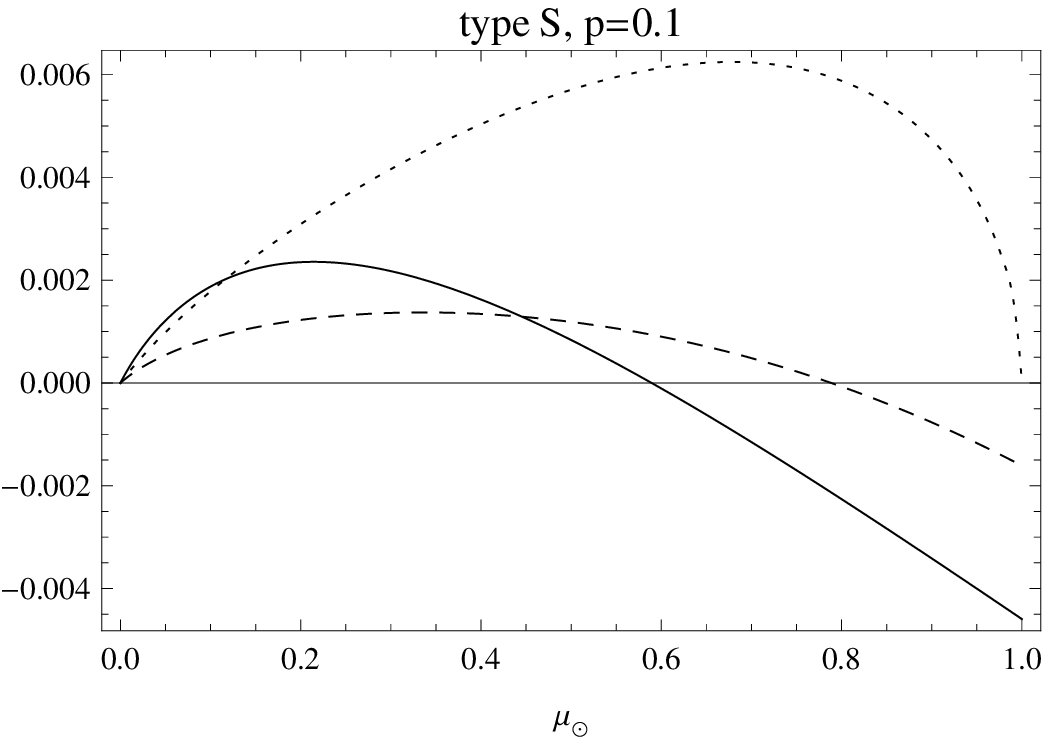}
  \end{center}
  \caption{Functions describing the non-Lambertian force model. Solid line represents the remainder $\mu_\odot (A_\mathrm{h} - A_\mathrm{B})$,
  dashed and dotted lines refer to $X_1$ and $X_2$ respectively.}
  \label{fig:1}
\end{figure}

The coefficient
\begin{eqnarray}
  \xi &=&  \frac{3}{2}\,\frac{I_3}{\epsilon_\mathrm{h}} - 1,
\end{eqnarray}
is a small quantity of order $10^{-2}$ or less. The two functions
\begin{eqnarray}
  X_1 &=&  \mu_\odot \left( \frac{A_\mathrm{h}\,I_3}{\epsilon_\mathrm{h}} - I_1 \right), \\
  X_2 &=& \mu_\odot\,I_2,
\end{eqnarray}
also represent a small deviation from the Lambertian model (see Fig.~\ref{fig:1} based upon the data from Tab.~\ref{tab:hap}).

In our judgement there is no point in producing excessively accurate approximations of corrections to the Lambertian model, so we use relatively simple
functions, found by trial and error,
\begin{eqnarray}
  X_1 & \approx &  \xi_{10}\,\mu_\odot \, \frac{ 1 - \mu_\odot - \xi_{11}\,\mu_\odot^2}{1 +3 \mu_\odot - \xi_{12}\,\mu_\odot^2} , \\
  X_2 & \approx & \xi_{20}\,\mu_\odot  \left( 3 - \xi_{21} \mu_\odot + \mu_\odot^2 \right)\,\sqrt{\frac{1-\mu_\odot}{1+\mu_\odot}},
\end{eqnarray}
with coefficients $\xi_i$  generated by the least squares adjustment to the results of numerical quadratures.

The limit case of Lambertian model results from setting $\xi = X_1 = X_2 = 0$, and then Eq.~(\ref{dF:fin}) simplifies to
\begin{equation}\label{dF:L}
    \frac{\mathrm{d}\vv{F}_\mathrm{L}}{\mathrm{d}S} =
   - \frac{2}{3c} \left(  \nu J\,\mu_\odot + Q \right) \vv{n}.
\end{equation}
Of course, this step also requires assuming a constant $A_\mathrm{h} = A_\mathrm{B}$ in boundary conditions (\ref{ener:1}) for a heat conduction
solver providing $Q$.

\subsection{Torque expression}

The force defined by Eq.~(\ref{dF:fin}) generates for each surface element a torque\footnote{We maintain the symbol $\vv{M}$ from previous papers hoping that it will not be confused with exitance $M$
appearing only in Sect.~\ref{scatra}.}
\begin{eqnarray}
  \mathrm{d}\vv{M} & = &  \left( \vv{r} \times \frac{\mathrm{d}\vv{F}}{\mathrm{d}S}\right)\,\mathrm{d}S =
  - \frac{2}{3}\,\frac{1+\xi}{c} \left[ \nu J \mu_\odot  + Q \right]\,\left(\vv{r} \times \mathrm{d}\vv{S} \right)  \nonumber \\
  &  & + \frac{\nu J}{c}\, \left[  X_1  \,\left(\vv{r} \times \mathrm{d}\vv{S} \right)
  -  X_2  \,\left( \vv{r} \times \vv{m}\, \mathrm{d}S \right) \right].  \label{dM}
\end{eqnarray}
The two cross products present in this formula differ in nature; the first, namely
\begin{equation}\label{rtS}
 \vv{r} \times \mathrm{d}\vv{S} = \vv{r} \times \vv{n}\,\mathrm{d}S,
\end{equation}
is constant over time in the body frame, whereas the second,
\begin{equation}\label{rtm}
    \vv{r} \times \vv{m}\, \mathrm{d}S  = \frac{ \vv{r} \times \vv{s} - \mu_\odot \,\vv{r}\times  \vv{n} }{s_\odot}\,\mathrm{d}S
    = \frac{\vv{r} \times \vv{s}}{s_\odot}\,\mathrm{d}S - \frac{\mu_\odot}{s_\odot}\,\left(\vv{r}\times \mathrm{d}\vv{S} \right),
\end{equation}
is time dependent due to the solar motion on local celestial sphere of $\mathrm{d}S$.

\begin{figure*}
 \begin{center}
  \includegraphics[width=16cm]{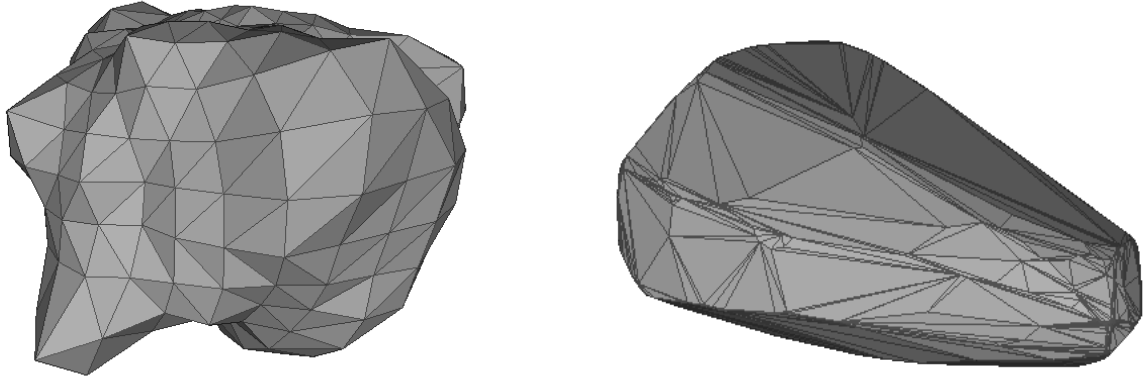}
  \end{center}
  \caption{Triangulated shape models: left -- (54509) YORP, right -- (3103) Eger.}
  \label{fig:mod}
\end{figure*}

\subsection{YORP effect computation}

Total YORP torque $\vv{M}$ resulting from the force (\ref{dF:fin})
\begin{equation}\label{M:tot}
    \vv{M} = \oint \left( \vv{r} \times \frac{\mathrm{d}\vv{F}}{\mathrm{d}S}\right)\,\mathrm{d}S,
\end{equation}
is obtained by integration over the body surface. The way the integration is handled depends on the type of a body shape model: it can be performed analytically
if the surface equation is explicit (e.g. spherical harmonics expansion) or, more often, replaced by
the sum over flat faces of a triangulation mesh. In the Rubincam approximation, when $Q=0$, one may simply substitute Eq.~(\ref{dF:fin}) into (\ref{M:tot})
to obtain the torque  for a given position of the Sun in the body frame. Most often the resulting torque values are then averaged with respect to rotation and
orbital motion in order to extract the secular effects in rotation rate and attitude dynamics. This step requires assumptions about the nominal
rotation model that provides the averaging kernel and solar ephemerides.

When the heat conduction is included, the nominal rotation model enters much earlier than in the final averaging:
surface temperature oscillations are lagged with respect to the insolation, hence we cannot find $Q$, required by the torque formula,
without the knowledge of rotation history. Choosing the simplest principal axis rotation mode (known as the gyroscopic approximation) we can easily add non-Lambertian
corrections to the algorithm of \citet{BBC:2010} based on a nonlinear 1D thermal model.

The one-dimensional model, where conduction is restricted to the direction normal to the surface, allows a separate treatment of each triangular face
of the shape mesh. There, having specified the obliquity $\varepsilon$ (angle between the spin axis and normal to the orbit)
we sample mean anomaly and rotation phase creating the vector of absorbed radiant flux values (the first term in the right-hand side of Eq.~(\ref{ener:2})).
Its discrete Fourier transform (DFT) serves to compute the DFT spectrum of $Q$ by an iterative process. Once the spectrum of $Q$ is known,
one is able to compute the torque $\vv{M}$. Until this step, no essential modifications of the algorithm are required; all one has to do
is to replace the constant albedo $A$ (understood as the Bond albedo $A_\mathrm{B}$) in the boundary conditions of \citet{BBC:2010} by the hemispheric albedo function
$A_\mathrm{h}(\mu_\odot)$. Apart from this point, the heat diffusion solver remains practically unchanged; however, computing the mean torque
demands a deeper revision. The spectrum of absorbed power flux $\nu\,J\,\mu_\odot (1-A_\mathrm{h})$, evaluated for the conductivity contribution,
cannot be recycled in the Rubincam part, because the hemispheric albedo is not a constant. Thus, outside the conductivity related block, we directly compute
the mean values of projections of the Rubincam part $\mathrm{d}\vv{M}$ (given by Eq.~(\ref{dM}) with $Q=0$) on unit vectors
\begin{equation}\label{vecs}
\begin{array}{l}
 \vv{e}_1  =
   \sin{\Omega'}\,\vv{ e}_x
   +\cos{\Omega'}\,\vv{ e}_y ,\\
 \vv{e}_2   =
   - \cos{\Omega'}\,\vv{ e}_x + \sin{\Omega'}\,\vv{ e}_y, \\
     \vv{e}_3  =  \vv{ e}_z, \\
\end{array}
\end{equation}
where $\vv{e}_x$, $\vv{e}_y$, $\vv{e}_z$ form the body-fixed frame basis, and $\Omega'$ is the rotation phase measured from the asteroid's equinox
\citep[prime added to avoid confusion with solid angle $\Omega$ of the present paper]{BBC:2010}.
Then we add the mean values resulting from the DFT spectrum of $Q$, obtaining the final averaged torque projections $\langle M_i \rangle = \langle \vv{M}\cdot \vv{e}_i \rangle$.

If $\omega$ stands for the rotation rate, the dynamics in the gyroscopic approximation is governed by
\begin{eqnarray}
  \dot{\omega} &=&  \frac{ M_3 }{C}, \label{eqm:1} \\
  \dot{\varepsilon} & = & \frac{ M_1}{
  \omega\,C}, \label{eqm:2} \\
  \dot{\Omega'} & = &
 \omega - \frac{M_2}{
  \omega\,C\,\tan{\varepsilon}}, \label{eqm:3}
\end{eqnarray}
where $C$ designates the maximum moment of inertia in the principal axes frame.

The conclusion of \citet{BBC:2010}, that all 1D thermal models imply the independence of the mean period related component $\langle M_3 \rangle$
on conductivity, holds true regardless of the scattering and emission laws.

\section{Exemplary results}

In order to see how the improvement of scattering and emission laws affects the simulated YORP effect, we considered two exemplary objects
out of the four known to have observationally confirmed spin acceleration: (54509) YORP asteroid with an irregular,
radar-determined shape model\footnote{More precisely, we use the `A-Rough' model available through the NASA PDS website.} \citep{Tay:07},
and (3103) Eger with a convex shape model obtained by lightcure inversion \citep{DVP:2009DPS}.
Both shape models, consisting of 572 (YORP) and 1972 (Eger) triangular facets are displayed in Fig.~\ref{fig:mod}.
Orbital and physical parameters assumed in our computations are presented in Table~\ref{params}. Generally, we tried to maintain
coherence with the data applied by \citet{Tay:07} and \citet{DVP:2009DPS}. The effective diameter (the radius of a sphere with
the same volume as an object) of Eger was selected indirectly: actually we scaled the asteroid to have the same volume as
a spheroid with semi-axes $2.3$ and $1.5$~km \citep{BOG:97}.
\begin{table}
\caption{Physical and orbital data for the test bodies}
~\\
\begin{tabular}{|l c|c|c|}
  \hline
  & & (54509) YORP  & (3103) Eger\\
  \hline
  epoch & JD & $2452117.5$  & $2446617.0$\\
  semi-axis & au & $0.9930$ & $1.4068$  \\
  eccentricity & & $0.2305$ & $0.3548$ \\
  inclination & deg & $1.9971$  & $20.939$ \\
  asc. node & deg & $283.835$ & $129.972$  \\
  arg. perihelion & deg & $272.091$ & $253.661$  \\
  \hline
  rotation period & h & $0.2029$ & $5.7102$  \\
  ecliptic pole $(\lambda,\beta)$ & deg & $(180,\,-85)$ & $(224,\,-72)$  \\
  \hline
  effective diameter & m & $113$ &  $1778$ \\
 density & $\mathrm{ kg\,m^{-3}}$ & $2500$ & $2800$ \\
 conductivity & $\mathrm{ W\,m^{-1} K^{-1}}$ & $0.02$ & $0.02$\\
 specific heat & $\mathrm{ J\,kg^{-1} K^{-1}}$ & $680$ & $680$\\
 max. mom. inertia & $\mathrm{kg\,m^2}$ & $3.04 \times 10^{12}$ & $2.80\times 10^{20}$ \\
 \hline
\end{tabular}
\label{params}
\end{table}

Considering the (54509) YORP, we compared two variants: a realistic assumption that the asteroid is an S-type object with
geometric albedo $p=0.1$, and rather ficticious case of spectral type E with $p=0.6$. The Hapke parameters were taken from
Table~\ref{tab:hap}. If the Lambert model was assumed, the appropriate Bond albedo and emissivity from Table~\ref{tab:hap}
were applied. For the Lommel-Seeliger model, the single scattering albedo $w$ was computed from $A_B$
according to Eq.~\ref{AB:LS} and the emissivity followed from $\epsilon_\mathrm{h} = 1-A_\mathrm{B}$.
Of course, the Lommel-Seeliger scattering was not considered for $p=0.6$  that might lead to
hemispheric albedo values outside the $[0,1]$ interval.

\begin{figure}
 \begin{center}
  \includegraphics[width=8cm]{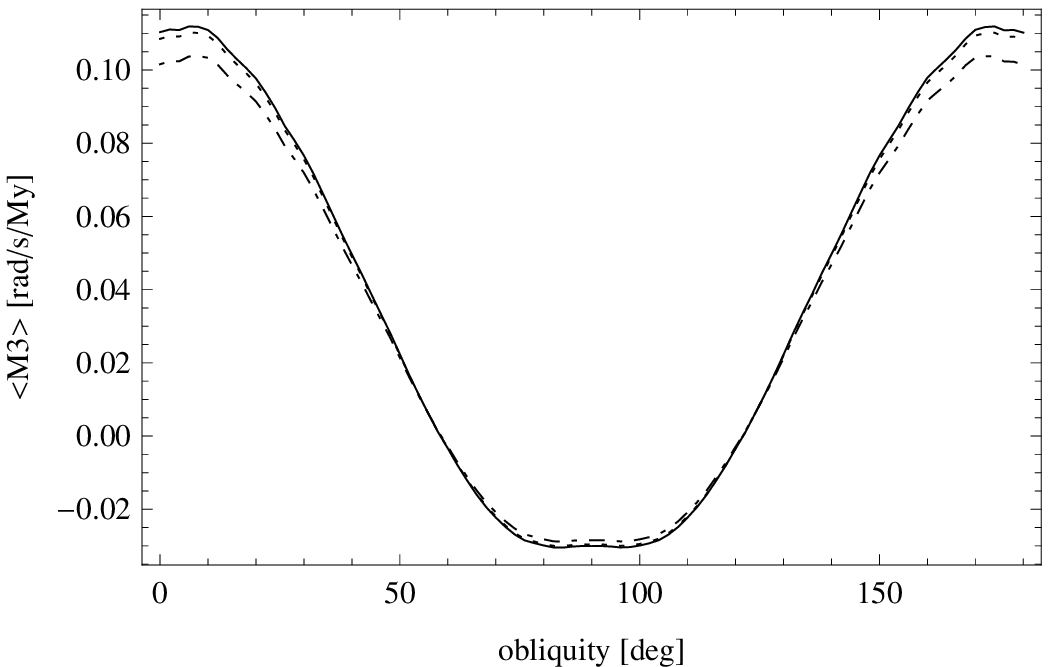}\\ \vskip0.5cm
  \includegraphics[width=8cm]{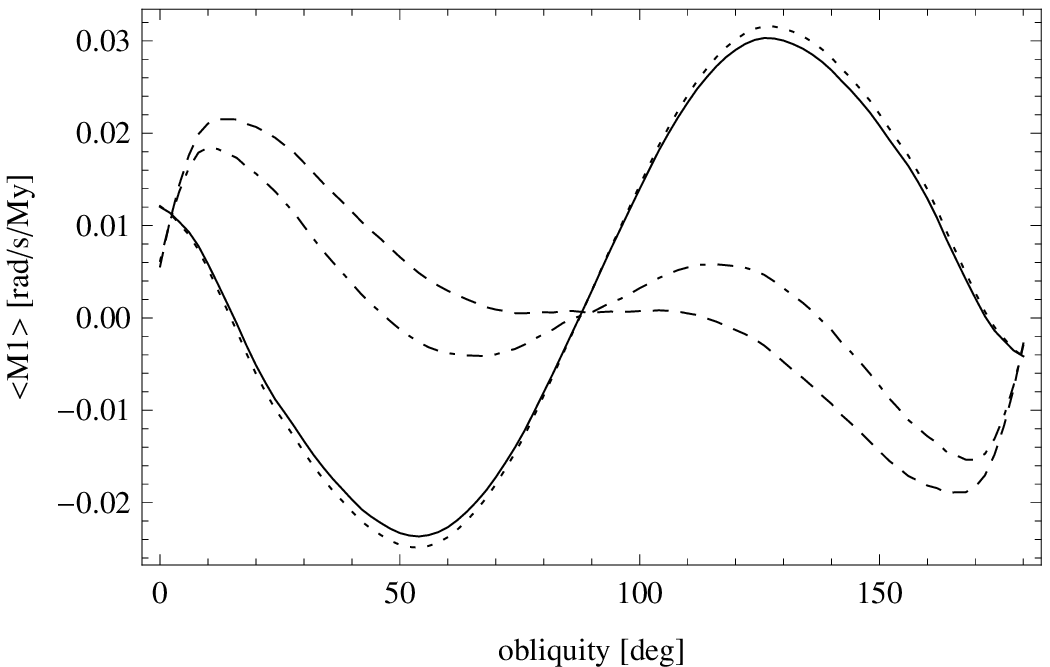}\\ \vskip0.5cm
  \includegraphics[width=8cm]{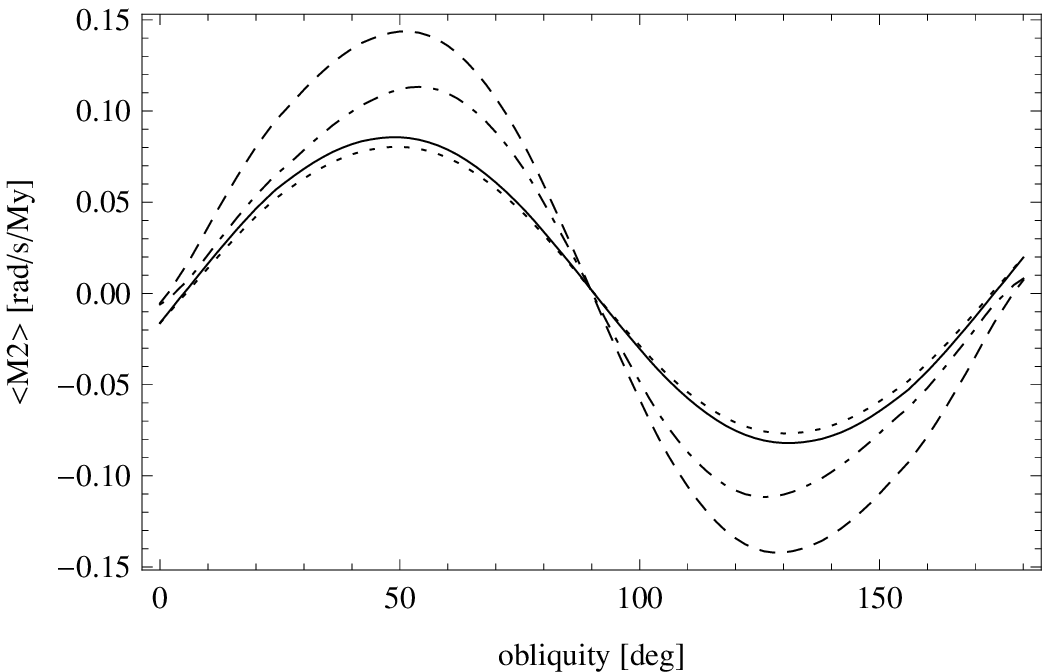}
  \end{center}
  \caption{Secular YORP effect components on (54509) YORP. Solid line -- Lambert type S , dotted -- Hapke type S ($p=0.1$), dashed -- Lambert type E,
  dot-dashed -- Hapke type E ($p=0.6$).}
  \label{fig:y}
\end{figure}

Figure~\ref{fig:y} presents the simulation results for (54509) YORP. Although its caption mentions only Lambert and Hapke models, the Lommel-Seeliger
results for $p=0.1$ are still there: they practically coincide with the Lambertian solid line. We traced the values of $\left\langle M_i \right\rangle$
for all possible obliquities $\varepsilon$, although the actual value for (5409) YORP is $\varepsilon = 173\degr$. The angle between
the asteroid's vernal equinox and the orbital perihelion $\omega_\mathrm{o}$, irrelevant for the rotation period related $\left\langle M_3 \right\rangle$, but
essential from the point of view of $\left\langle M_1 \right\rangle$ and $\left\langle M_2 \right\rangle$, responsible for the attitude \citep{BBC:2010},
is $\omega_\mathrm{o} = 102\degr$ and we used this value for all $\varepsilon$ values in Fig.~\ref{fig:y}.
Considering $\left\langle M_3 \right\rangle$ values (Fig.~\ref{fig:y}, top),  we observe that in spite of irregular shape the kind of scattering model
at low $p=0.1$ has practically no influence on the YORP effect in rotation period, and even at the high albedo case ($p=0.6$)
the difference between Lambert and Hapke models does not exceed 10 percent.
The situation is different for $\left\langle M_1 \right\rangle$ and
$\left\langle M_2 \right\rangle$, but there, even for the Lambert model,  we have a dependence on albedo resulting from the heat conduction.
Although for $p=0.1$ there is almost no difference between the Lambert, Lommel-Seeliger, and type S Hapke models, a high geometric albedo $p=0.6$
leads to significant differences between the Lambert approximation (dashed) and the E-type Hapke model (dot-dashed).
The results for nominal values of $\varepsilon = 173\degr$ and $\omega_\mathrm{o}=102\degr$ are collected in Tab.~\ref{wyn:y}. Comparing
$\left\langle M_3 \right\rangle$ with the observed $\dot{\omega} = (4.7 \pm 0.5) \times 10^{-16}~\mathrm{rad\,s^{-2}}$ \citep{Tay:07}, we note that
the present model overestimates $\dot{\omega}$ almost $7.6$ times, i.e. more than the relevant models used in \citet{Tay:07}. On the other hand,
the most significant part of this increase is due to the recomputed reduction to the centre of mass and (more important) principal axes system.
If the original body fixed frame is used, we obtain a lower factor $7.0$.

\begin{figure}
 \begin{center}
  \includegraphics[width=7.75cm]{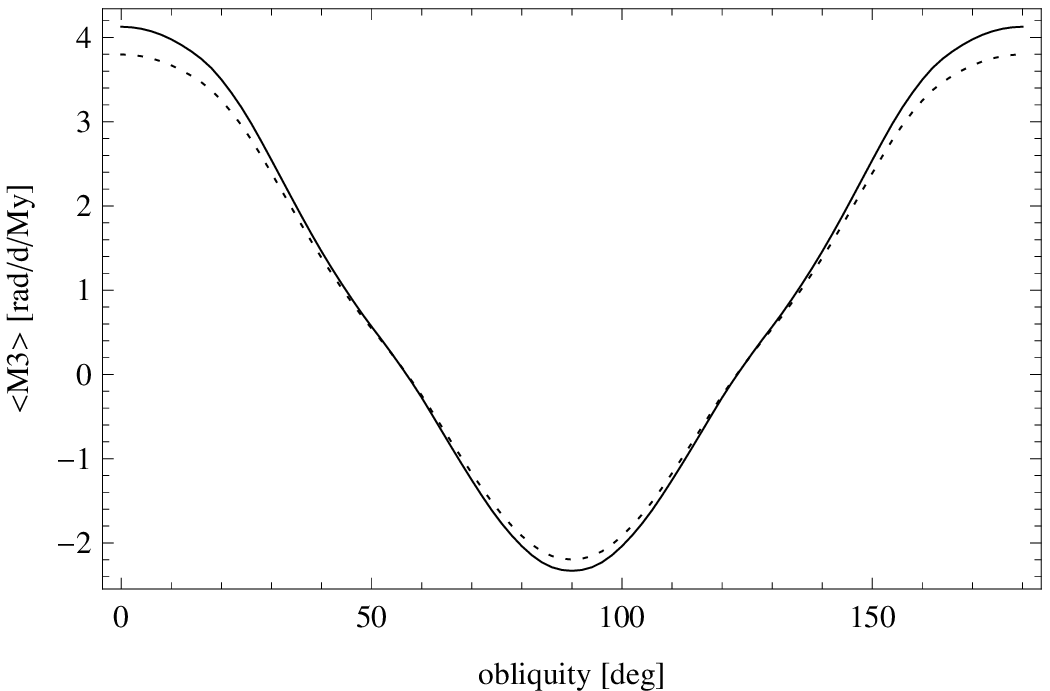}\\ \vskip0.5cm
  \includegraphics[width=7.75cm]{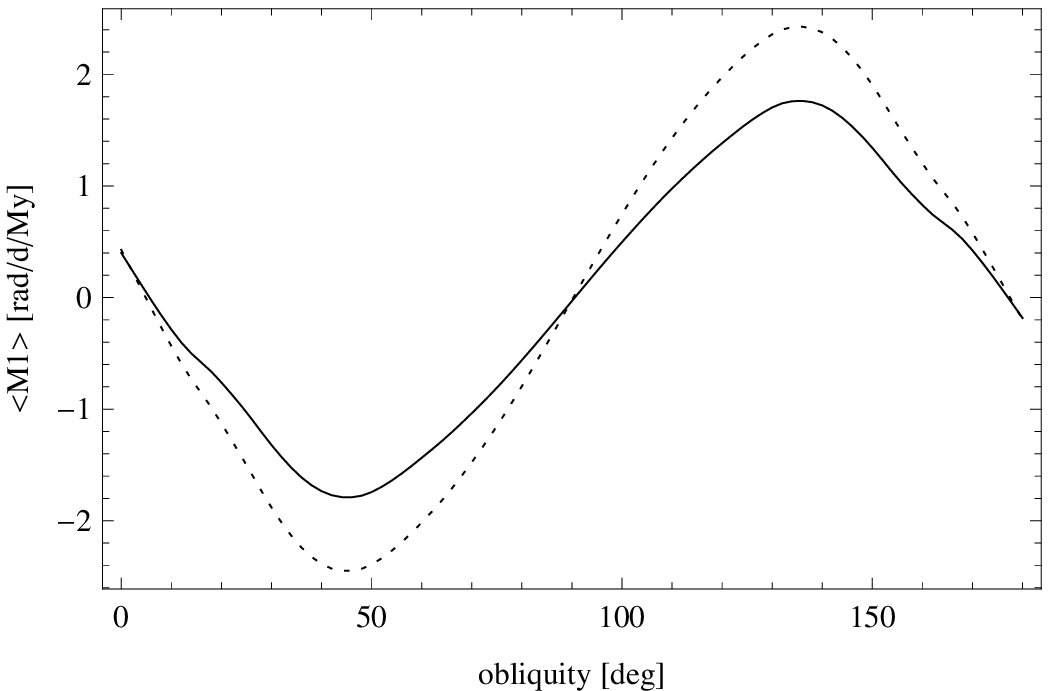}\\ \vskip0.5cm
  \includegraphics[width=7.75cm]{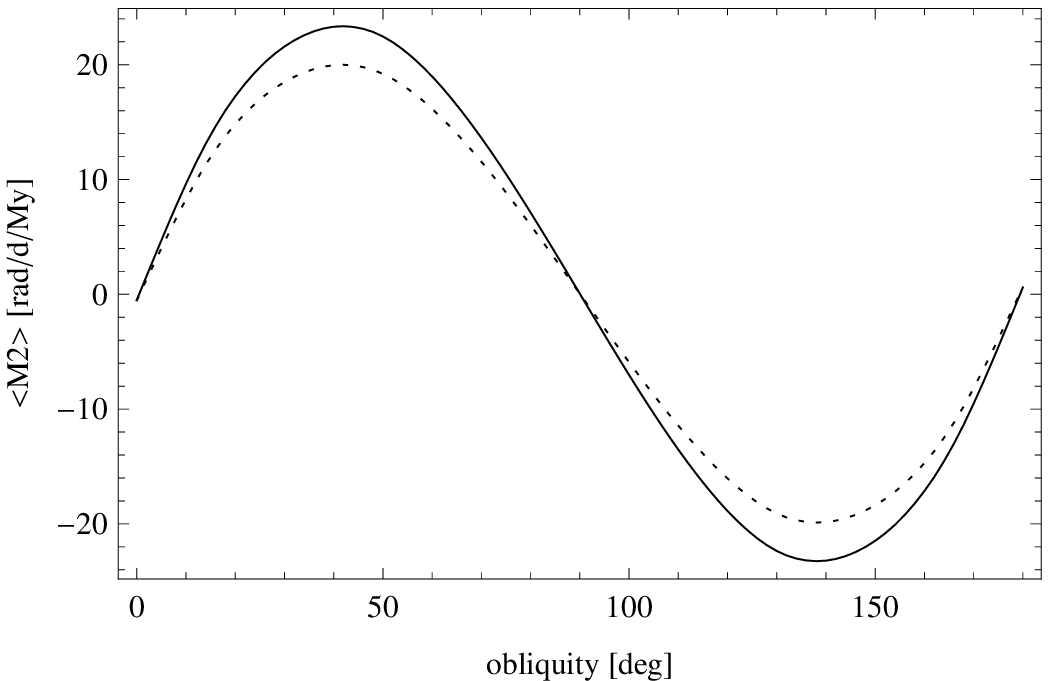}
  \end{center}
  \caption{Secular YORP effect components on (3103) Eger. Solid line -- Lambert type E , dotted -- Hapke type E ($p=0.6$).}
  \label{fig:e}
\end{figure}

\begin{table}
\caption{Mean YORP torques for (54509) YORP at $\varepsilon = 173\degr$ and $\omega_\mathrm{o}=102\degr$.
All values in $10^{-16}~\mathrm{rad\,s^{-2}}$.}
~\\
\begin{tabular}{|l c|c|c|c|}
  \hline
    &   & $\left\langle M_1 \right\rangle$ & $\left\langle M_2 \right\rangle$ & $\left\langle M_3 \right\rangle$ \\
  \hline
  Lambert & $p=0.1$ & $-0.21$  & $-0.92$ & $35.5$ \\
  Lommel-Seeliger & $p=0.1$ & $-0.18$ & $-0.46$ & $35.4$  \\
  Hapke (type S) & $p=0.1$ & $-0.07$ & $-0.34$ & $34.9$ \\
 \hline
  Lambert & $p=0.6$ & $-5.07$  & $-6.41$ & $35.5$ \\
  Hapke (type E) & $p=0.6$ & $-4.44$ & $-2.82$ & $32.9$ \\
  \hline
\end{tabular}
\label{wyn:y}
\end{table}

In the simulations referring to (3103) Eger we compared only the Lambert and Hapke model for spectral type E with a high geometric albedo
$p=0.6$. In spite of a convex shape, excluding all shadowing effects, the dependence of all three $\left\langle M_i \right\rangle$ components
on the scattering/emission model has the same relative magnitude as in the case of (54509) YORP. The values for the actual spin axis orientation
of Eger are provided in Tab.~\ref{wyn:e}. Interestingly, our modeled values of $\left\langle M_3 \right\rangle$ are very close to the
observed $\dot{\omega} = (1.2 \pm 0.8) \times 10^{-18}~\mathrm{rad\,s^{-2}}$ reported by \citet{DVP:2009DPS}. Of course, this exceptional agreement can be a lucky coincidence,
recalling inaccurate nature of photometric convex shape model, roughly estimated density and still a large error margin of the
$\dot{\omega}$ determination.

\begin{figure}
 \begin{center}
  \includegraphics[width=8cm]{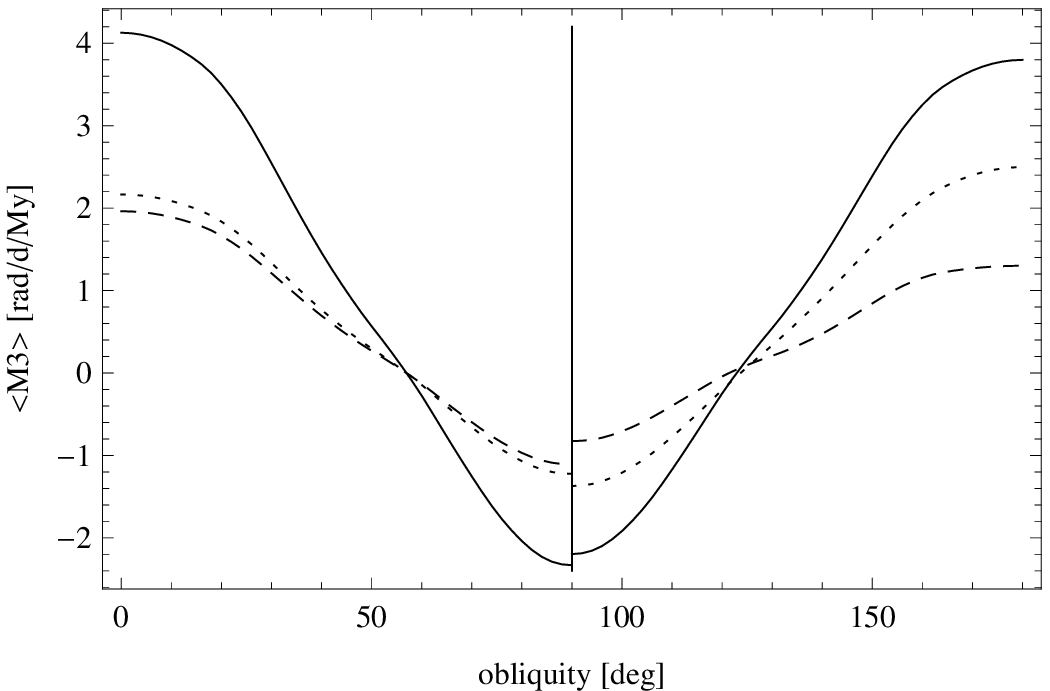}
  \end{center}
  \caption{YORP effect in $\omega$ and its components for (3103) Eger. Dashed line -- scattered light, dotted -- thermal radiation,
  solid line -- total. Left half shows the Lambert case compared with the Hapke model (right half).}
  \label{fig:ecm}
\end{figure}

\begin{table}
\caption{Mean YORP torques for (3103) Eger at $\varepsilon = 177\degr$ and $\omega_\mathrm{o}=100\degr$.
All values in $10^{-18}~\mathrm{rad\,s^{-2}}$.}
~\\
\begin{tabular}{|l c|c|c|c|}
  \hline
    &   & $\left\langle M_1 \right\rangle$ & $\left\langle M_2 \right\rangle$ & $\left\langle M_3 \right\rangle$ \\
 \hline
  Lambert & $p=0.6$ & $0.002$  & $-0.93$ & $1.51$ \\
  Hapke (type E) & $p=0.6$ & $0.015$ & $-0.77$ & $1.39$ \\
  \hline
\end{tabular}
\label{wyn:e}
\end{table}

\section{Conclusions}

As far as photometry of Solar System bodies is concerned,  the bidirectional reflectance model elaborated by Hapke leads to
significantly different results than the basic Lambertian framework. But the YORP effect in rotation period occurs to be
almost insensitive to the scattering/emission model and even at highest observed albedo values the difference between
the two models does not increase to more than 10 percent. However, this low sensitivity should not be interpreted as
the evidence of insensitivity of the scattered radiation torque on reflectivity model. Actually, the situation
quite opposite. Even for a given Bond albedo value, the part of the YORP torque originating from the recoil of reflected
light significantly depends on the form of BRDF. But the conservation of energy implies that in the absence of conductivity
the sum of scattered and thermally re-radiated energy is always equal to the incident energy. If the hemispheric albedo
in some reflection model is higher than the Bond albedo of the Lambert case, more power is scattered, but also less power is
thermally re-emitted and vice versa (see Fig.~\ref{fig:ecm}). Actually, the same mechanism of energy balance is responsible the independence of YORP
on albedo and emissivity in the traditional Rubincam approximation with Lambertian scattering/emission.
In the 1D thermal model considered in our paper the $\left\langle M_3 \right\rangle$ component behaves exactly like in the
Rubincam's approximation, so the dependence on reflectance is only due to secondary effects -- mostly related with the small deviation
of the recoil force from the normal to the surface.

Using a more elaborate scattering method is more important in the part of the YORP effect responsible for the orientation of the spin axis.
In Rubincam approximation, the situation is similar to that of $\left\langle M_3 \right\rangle$. But the Rubincam approximation itself is
definitely unrealistic for the attitude even at moderate values of conductivity. The influence of heat conduction is proportional to the
absorbed fraction of incident energy, hence to the albedo. It means that two scattering models with different dependence of hemispheric albedo
on Sun zenith distance will differently affect the balance between scattered and re-radiated power. This explains why using the Hapke BRDF
instead of Lambertian model is more important for $\left\langle M_1 \right\rangle$ and $\left\langle M_2 \right\rangle$, then for
$\left\langle M_3 \right\rangle$.

\section*{Acknowledgments}

The work of S. Breiter was supported by the Polish Ministry of Science and Higher Education -- grant N N203 302535.
The work of D. Vokrouhlick\'{y} was supported by the Czech Grant Agency (grant 205/08/0064) and the Research Program
MSM 0021620860 of the Czech Ministry of Education.


\appendix

\section{Lommel-Seeliger approximation}
\label{app:a}

The Lommel-Seeliger scattering law \citep{Fairb:2005} is defined by the BRDF
\begin{equation}\label{BDRF:LS}
    f_\mathrm{LS}(\vv{s},\vv{o}) =   \frac{w}{4\pi\,(\mu_\odot+\mu)},
\end{equation}
which can be seen as a simplified Hapke model independent on the phase angle $g$.
Using this simple law we find most of the expressions in exact, closed form, depending on the single scattering albedo $w$.
The hemispheric albedo is
\begin{equation}\label{Ah:LS}
    A_\mathrm{h}(\mu_\odot) = \frac{w}{2}\,\left( 1 + \mu_\odot \ln{\frac{\mu_\odot}{1+\mu_\odot}}\right),
\end{equation}
leading to the Bond albedo
\begin{equation}\label{AB:LS}
    A_\mathrm{B} = \frac{2\,(1-\ln 2)}{3}\,w \approx 0.204569\,w,
\end{equation}
and the geometric albedo
\begin{equation}\label{AG:LS}
    p = \frac{w}{8}.
\end{equation}
Note that the last formula leads to problems with $w > 1$ in Hapke's thermal radiation expressions if we try to use it for bright objects with $p > 0.125$.
In these circumstances we combine the Lommel-Seeliger scattering with a Lambertian grey body emission model, imposing $\xi=0$. Then,
\begin{equation}\label{X1:LS}
    X_1(\mu_\odot) = \frac{w\,\mu_\odot}{12} \,\left[1 + 6 \mu_\odot + 2 \mu_\odot (2 + 3 \mu_\odot) \ln\frac{\mu_\odot}{1 + \mu_\odot}\right],
\end{equation}
and $X_2 = 0$. The resulting force per area
\begin{equation}\label{dF:LS}
  \frac{\mathrm{d}\vv{F}}{\mathrm{d}S} = - \frac{2}{3c}  \left[ \nu J \mu_\odot  + Q \right]\,\vv{n}
    + \frac{\nu J}{c}\,   X_1  \,\vv{n},
\end{equation}
is directed along the surface normal, similarly to the Lambertian case.

\section{Direct radiation pressure}

In the usual YORP models of a single, Sun orbiting object the direct radiation pressure, opposite to the Sun vector $\vv{s}$,
is either a priori discarded  or its effect disappears after the double (rotation and orbit) averaging. However, this phenomenon
may play some role when a binary system is studied -- most notably for the BYORP effect \citep{CB:2005,MSch:2010}. In such cases,
the force and torque should be complemented with following terms:
\begin{equation}\label{Frp}
  \frac{\mathrm{d}\vv{F}_\mathrm{d}}{\mathrm{d}S} = - \frac{ \nu\,J \mu_\odot }{c}  \,\vv{s}  =
  - \frac{ \nu\,J \mu_\odot }{c}  \,\left( \mu_\odot \vv{n} + s_\odot \vv{m} \right),
\end{equation}
being the addition to Eq.~(\ref{dF:fin}), and the resulting torque
\begin{equation}\label{Mrp}
  \frac{\mathrm{d}\vv{M}_\mathrm{d}}{\mathrm{d}S} = \vv{r} \times \frac{\mathrm{d}\vv{F}_\mathrm{d}}{\mathrm{d}S}.
\end{equation}
Note that for a binary system the visibility function $\nu$ additionally involves occlusions by the second object.

With these complements, the complete force $\mathrm{d}\vv{F}_\mathrm{c}= \mathrm{d}\vv{F}+\mathrm{d}\vv{F}_\mathrm{d}$
acting on a surface element is
\begin{eqnarray}
  \mathrm{d}\vv{F}_\mathrm{c}  &  = & - \frac{2}{3}\,\frac{1+\xi}{c} \left[ \nu J \mu_\odot    + Q \right]\,\mathrm{d}\vv{S}
  \nonumber \\
   & &  + \frac{\nu J}{c}\, \left[  \left( X_1 - \mu_\odot^2 \right) \,\mathrm{d}\vv{S}
  - \left( X_2 + \mu_\odot\,s_\odot \right)  \,\vv{m} \mathrm{d}S \right], \label{dF:compl}
\end{eqnarray}
and the complete torque is readily obtained by the cross product
\begin{equation}
\mathrm{d}\vv{M}_\mathrm{c} = \vv{r} \times \mathrm{d}\vv{F}_\mathrm{c}.
\end{equation}

Equation~(\ref{Frp}) involves an implicit statement that all photons hitting the surface are absorbed and transfer their momentum
before being re-emitted in any form, including the one termed reflection. In a perfect specular reflection, all photons arriving from
$\vv{s} = \mu_\odot \vv{n} + s_\odot \vv{m},$ leave the surface in the symmetric direction $\vv{s}' = \mu_\odot \vv{n} - s_\odot \vv{m}$.
So, the total effect of perfect specular reflection is
\begin{equation}\label{Fs}
   \mathrm{d}\vv{F}_\mathrm{s}   =
  - \frac{2\, \nu\,J \mu_\odot^2 }{c}  \, \mathrm{d}\vv{S},
\end{equation}
with the $\vv{m}$ component canceled. As a consequence, we may interpret the occurrence of function $X_2$ as a footprint of imperfect
specular reflection in the Hapke's model, with only a part of power incoming from $\vv{s}$ leaving the surface along $\vv{s}'$.
\label{lastpage}
\end{document}